\documentclass[prb,letterpaper,twocolumn,showpacs,superscriptaddress,amsmath,amsfonts,amssymb,preprintnumbers,balancelastpage,citeautoscript,byrevtex]{revtex4}
\pdfoutput=1
\usepackage[T1]{fontenc}\usepackage[latin1]{inputenc}
\usepackage{dcolumn,graphicx,color,booktabs,microtype} %\graphicspath{{Figures/}}
\usepackage[charter]{mathdesign}
\renewcommand{\figurename}{Fig.}
\renewcommand{\tablename}{Table}
\makeatletter\renewcommand{\fnum@figure}[1]{\figurename~\thefigure.}\makeatother
\makeatletter\renewcommand{\fnum@table}[1]{\tablename~\thetable.}\makeatother
\definecolor{Red}{rgb}{0.8,0,0.0}

\usepackage[colorlinks,plainpages=false,linkcolor=black,urlcolor=blue,citecolor=black,pdfpagemode=UseNone,pdfstartview=FitBH]{hyperref}

\begin{document} \pagestyle{plain}

\title{Temperature-dependent Fermi surface of 2H-TaSe$_2$ \\driven by competing density wave order fluctuations}

\author{D.\,S.\,Inosov}
\affiliation{Institute for Solid State Research, IFW Dresden, P.\,O.\,Box 270116, D-01171 Dresden, Germany.}
\affiliation{Max-Planck-Institut für Festkörperforschung, Heisenbergstraße 1, 70569 Stuttgart, Germany.}
\author{D.\,V.~Evtushinsky}\author{V.\,B.\,Zabolotnyy}
\affiliation{Institute for Solid State Research, IFW Dresden, P.\,O.\,Box 270116, D-01171 Dresden, Germany.}
\author{A.\,A.\,Kordyuk}
\affiliation{Institute for Solid State Research, IFW Dresden, P.\,O.\,Box 270116, D-01171 Dresden, Germany.}
\affiliation{Institute of Metal Physics of National Academy of Sciences of Ukraine, 03142 Kyiv, Ukraine.}
\author{B.\,Büchner}
\affiliation{Institute for Solid State Research, IFW Dresden, P.\,O.\,Box 270116, D-01171 Dresden, Germany.}
\author{R.\,Follath}
\affiliation{BESSY GmbH, Albert-Einstein-Strasse 15, 12489 Berlin, Germany.}
\author{H.\,Berger}
\affiliation{Institut de Physique de la Mati\`{e}re Complexe, EPFL, 1015 Lausanne, Switzerland.}
\author{S.\,V.~Borisenko}
\affiliation{Institute for Solid State Research, IFW Dresden, P.\,O.\,Box 270116, D-01171 Dresden, Germany.}



%\preprint{\textit{Preprint: \today}}

\begin{abstract}

\noindent Temperature evolution of the 2H-TaSe$_2$ Fermi surface (FS) is studied by high-resolution angle-resolved
photoemission spectroscopy (ARPES). High-accuracy determination of the FS geometry was possible after measuring electron
momenta and velocities along all high-symmetry directions as a function of temperature with subsequent fitting to a
tight-binding model. The estimated incommensurability parameter of the nesting vector agrees with that of the
incommensurate charge modulations. We observe detectable nonmonotonic temperature dependence of the FS shape, which we
show to be consistent with the analogous behavior of the pseudogap. These changes in the electronic structure could stem
from the competition of commensurate and incommensurate charge density wave order fluctuations, explaining the puzzling
reopening of the pseudogap observed in the normal state of both transition metal dichalcogenides and high-$T_{\rm c}$
cuprates.

\end{abstract}

%\preprint{\textit{This preprint is available at arXiv:XXXX.XXXX.}}
%\preprint{\textit{To appear in the New Journal of Physics.}}

\maketitle

\vspace{-5pt}\section{Introduction}\vspace{-5pt}

\noindent 2H-TaSe$_2$ (trigonal prismatic tantalum diselenide) is a quasi-two-dimensional metal with layered hexagonal
crystal structure, which is remarkable for its two well-pronounced charge density wave (CDW) transitions at accessible
temperatures: an incommensurate charge density wave (ICDW) transition at $T_{\rm ICDW}=$122.3\,K and a commensurate
three-fold CDW lock-in transition at $T_{\rm CCDW}=$90\,K \cite{WilsonDiSalvo75, MonctonAxe75}. Such wealth of the phase
diagram combined with perfect suitability for surface-sensitive measurements make this material an ideal starting point
for studying complicated charge ordering phenomena in two-dimensional metals, including high-temperature superconductors
\cite{ReviewsChargeOrder}.

The Fermi surface (FS) is well-defined in its usual sense only in perfectly periodic crystals \cite{AshcroftMermin}. Indeed, if the potential loses its periodicity (e.g. due to the appearence of the incommensurate component), the quasimomentum is no longer a good quantum number, and the definitions have to be generalized. As long as the incommensurate component of the potential is weak, the spectral weight at the Fermi level remains localized near the original FS of the parent compound. This lets us generalize the definition of the FS as the locus of quasimomenta at the Fermi level that support the maximal spectral weight. Consequently, even if the spectral weight is reduced due to the opening of the pseudogap (PG), the ``remnant'' FS can still be defined. Throughout this paper, we will follow this generalized definition, unless otherwise specified.

\begin{figure}[b]
        \includegraphics[width=\columnwidth]{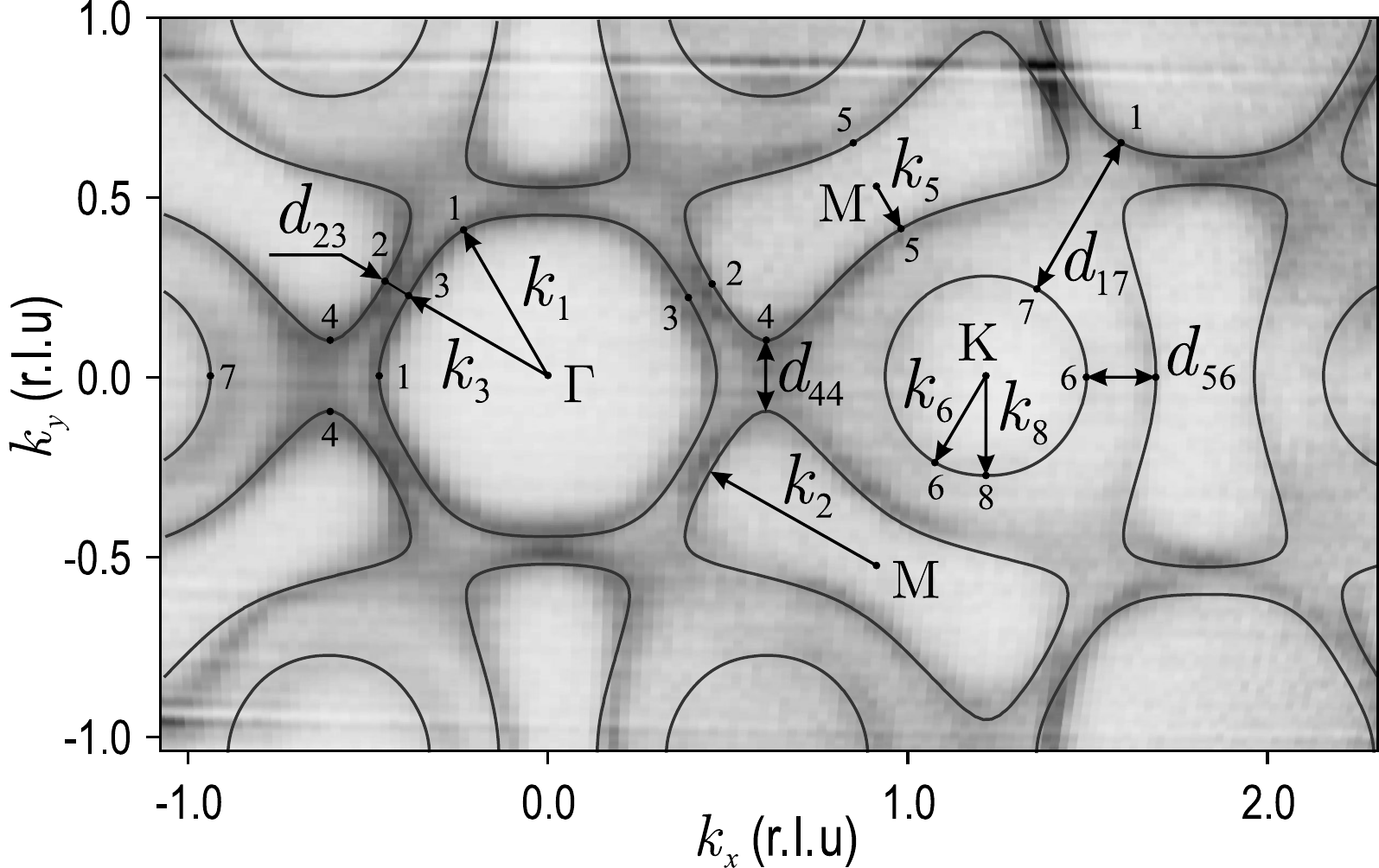}
        \caption{Normal-state Fermi surface of 2H-TaSe$_2$ at 276\,K (darker regions correspond to higher
        photoemission intensity). The contours represent the tight-binding fit to the experimental
        data. This figure serves as a legend for Fig.\,\ref{Fig:TdepFS}, defining the positions in
        $\mathbf{k}$-space where Fermi momenta and velocities along the high-symmetry directions have been measured as
        a function of temperature. The momentum scales are presented in relative lattice units (r.\,l.\,u.).\vspace{-4pt}}\label{Fig:Map276K}
\end{figure}

As we have previously observed \cite{BorisenkoKordyuk08}, the FS of 2H-TaSe$_2$ undergoes a 3$\times$3 reconstruction at
the commensurate CDW (CCDW) transition. The high-temperature FS measured by angle-resolved photoemission spectroscopy
(ARPES) is shown in Fig.\,\ref{Fig:Map276K}. The FS sheets shown by solid grey lines originate from two bands: one is
responsible for the $\mathrm{\Gamma}$ and K barrels with a saddle point in between, the other one supports the
``dogbone'' with one saddle point at the M point below the Fermi level and another one in between the ``dogbones'' above
the Fermi level. It was shown that upon cooling below the ICDW transition, the spectral weight at the Fermi level is
depleted at some $\mathbf{k}$-points due to the opening of the momentum-dependent PG \cite{BorisenkoKordyuk08}, which
gradually evolves into the band gap at $T_{\rm CCDW}$, culminating in the change of the FS topology. Below $T_{\rm
CCDW}$, the folded FS consists of a set of nearly circles around the new $\mathrm{\Gamma}$ points and rounded triangles
around new $\mathrm{K}$ points \cite{BorisenkoKordyuk08}. The gradual evolution of the FS from the high-temperature
(``normal'') state across the ICDW transition into the incommensurate state is the topic of the present paper.

\vspace{-5pt}\section{Pseudogap and its relation to the Fermi surface geometry}\vspace{-5pt}

The appearence of the PG in the incommensurate state is not surprising, as the depletion of the spectral weight near the
Fermi level has long been theoretically predicted in different nonperiodic systems (either quasicrystals or ICDW
compounds), independent of the nature of nonperiodicity \cite{PseudogapCollection}. A less trivial observation is that
the FS gradually changes its shape both above and below the ICDW transition temperature, as was first observed by
measuring the distance between the K and M barrels in 2H-TaSe$_2$ as a function of temperature
\cite{BorisenkoKordyuk08}. This change is nonmonotonic, which suggests its common origin with the puzzling reopening of
the PG above $T_{\rm ICDW}$ \cite{BorisenkoKordyuk08, BorisenkoNbSe2}---\,an effect also observed in high-$T_{\rm c}$
cuprates \cite{KordyukBorisenko08}. In earlier studies of 2H-TaSe$_2$, the \mbox{minor} temperature variations of the FS
could not be detected, and the dispersion was considered unchanged down to 90\,K \cite{UnchangedCollection}, until the
FS reconstruction due to the CCDW lock-in transition finally occurred. On the other hand, temperature dependence of the
FS in other ICDW-bearing materials, such as the quasi-one-dimensional K$_{0.3}$MoO$_3$ \cite{KMoO3Collection}, has been
known since a few years.

\begin{figure}[b]
        \includegraphics[width=\columnwidth]{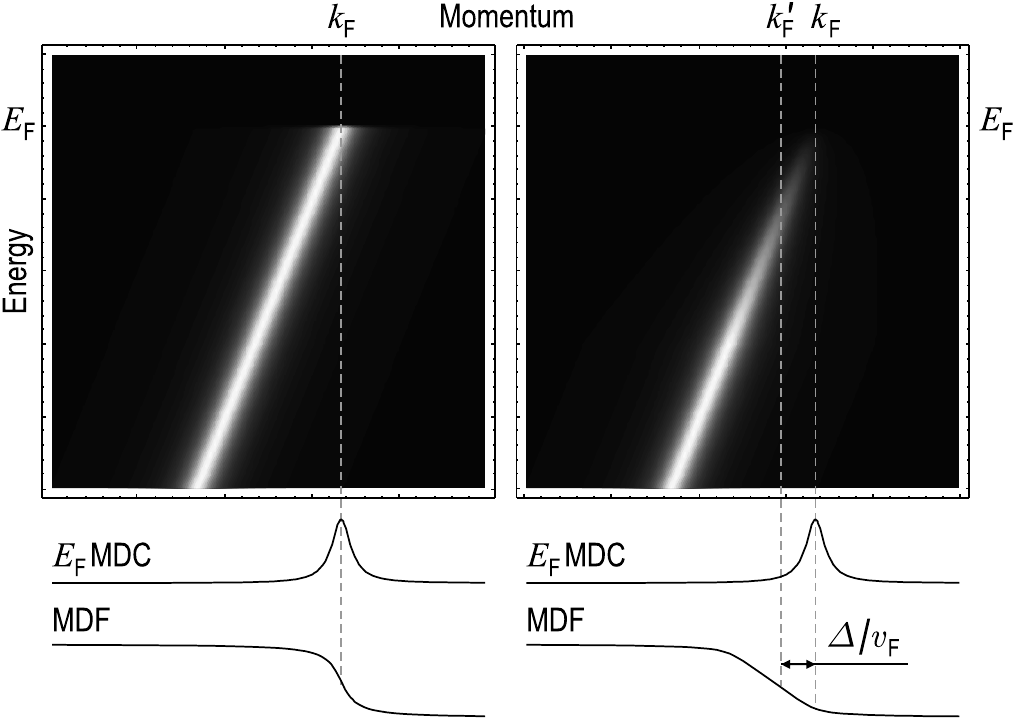}
        \caption{Model of the spectral function with linear dispersion and constant scattering rate in the vicinity of the Fermi level $E_{\rm F}$ without the PG (left) and with a finite PG (right). The black and white colors correspond to the zero and maximal spectral weight respectively. The curves at the bottom of both panels show the corresponding momentum distribution curves at the Fermi level ($E_{\rm F}$\,MDC) and the MDF. One can see that in the absence of the PG the peak position in the $E_{\rm F}$\,MDC coincides with the position of the step in the MDF, uniquely defining the value of the Fermi momentum $k_{\rm F}$. In the presence of the PG, these values differ by approximately $\Delta/v_{\rm F}$, which makes the definition of the Fermi momentum, and therefore the FS itself, ambiguous.\vspace{-4pt}}\label{Fig:Cartoon}
\end{figure}

To understand why the opening of the PG may have an effect on the shape of FS contours, we recall that the number of
occupied states in the conduction band, which can be calculated by integrating the momentum distribution function (MDF)
over the whole Brillouin zone, should be equal to the chemical doping level of the material \cite{Luttinger60}. In the
absence of the PG, the MDF is constant within every FS contour, and therefore for a two-dimensional metal the integral
is simply given by the FS area. As the PG opens, it leads to a depletion of the spectral weight at the FS, as illustrated in Fig.\,\ref{Fig:Cartoon}. Charge conservation requires that the total number of occupied states in
the conduction band remains constant, which means that the mid-point of the MDF onset, marked as $k'_{\rm F}$ in the right panel, should remain on average unchanged. This point deviates by approximately $\Delta/v_{\rm F}$ from the Fermi momentum $k_{\rm F}$, if one defines the latter naturally as the maximum of the momentum distribution curve (MDC) at the Fermi level $E_{\rm F}$. Here $\Delta$ is the PG, and $v_{\rm F}$ is the Fermi velocity. It is therefore clear that for the charge conservation to hold, the remnant FS has to adjust its shape and volume in consistence with the opening of the PG, causing an effective change of the area of each remnant FS barrel of the order of $\oint\!\Delta(\mathbf{k})/v_{\rm F}(\mathbf{k})\,\mathrm{d}k$. Here the momentum-dependent pseudogap $\Delta(\mathbf{k})$ and Fermi velocity $v_{\rm F}(\mathbf{k})$ are assumed, and the contour integration is performed along the FS. Such a change may lead to the spurious breakdown of the Luttinger theorem for the remnant FS and to the change of the
nesting conditions, which are determined by the Fermi momenta $k_{\rm F}$, rather than $k'_{\rm F}$. In other words, in the PG phase the definition of the FS becomes ambiguous, and one should clearly distinguish the ``remnant FS'' given by $k_{\rm F}$, which determines the nesting properties and is usually determined in ARPES works, from the ``charge FS'' given by $k'_{\rm F}$, which should conserve the FS area in accordance with the Luttinger theorem.

In the following, we demonstrate that this subtle effect of the PG can be measured experimentally by a careful examination of the temperature dependence of the remnant FS, providing a useful consistency check for the directly measured PG evolution with temperature. We would like to emphasize that the direct measurements of the pseudogap are usually based on the leading edge shift in the energy distribution curves (EDC), whereas the effects reported in the present paper follow from MDC analysis only. Consistency of both methods therefore indicates that the observed temperature dependence of the PG can not be attributed to an artifact of the method.

\vspace{-5pt}\section{Experimental observations}\vspace{-5pt}

\makeatletter\renewcommand{\fnum@figure}[1]{\figurename~\thefigure~(color online).}\makeatother
\begin{figure*}[t]
        \includegraphics[width=\textwidth]{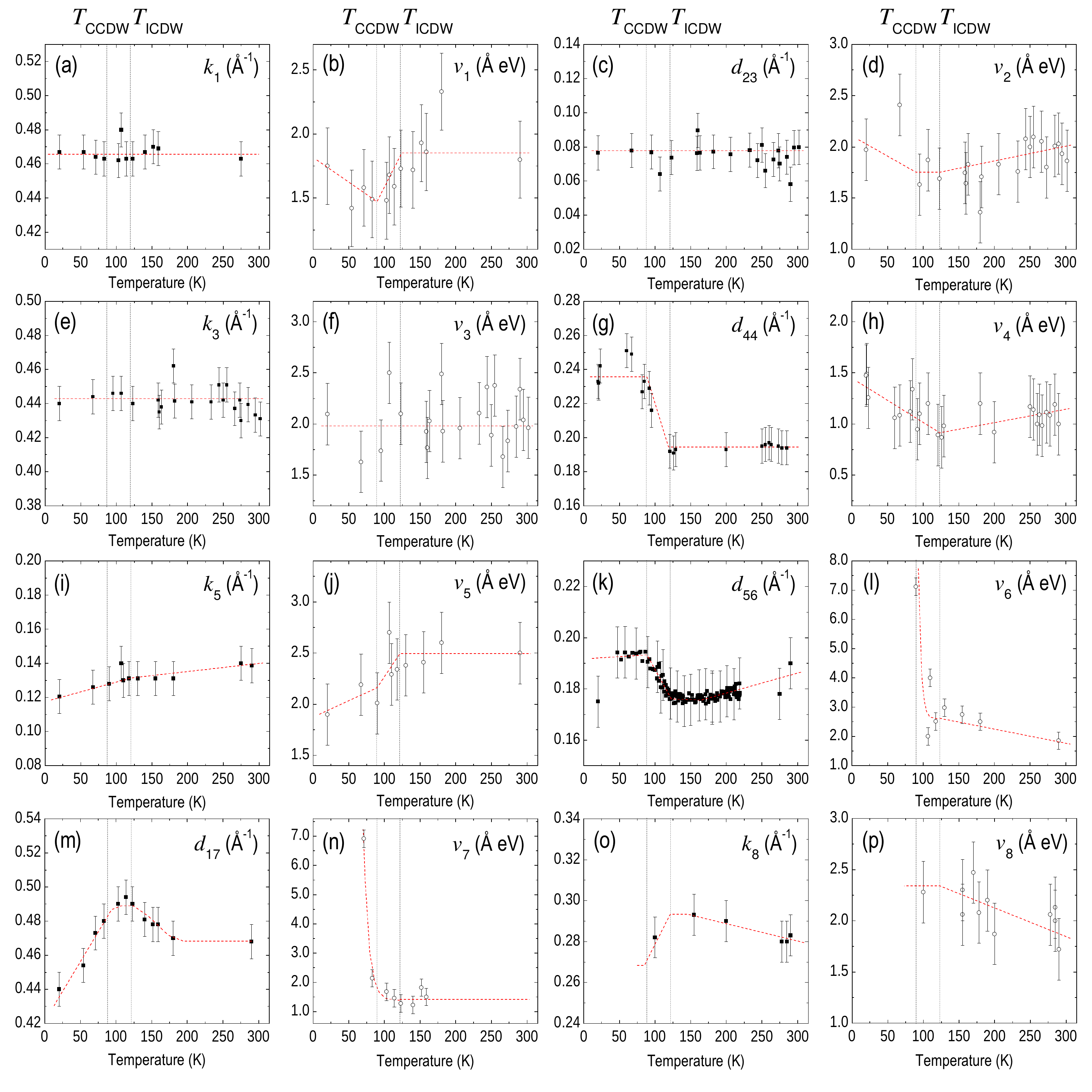}\vspace{-0.5em}
        \caption{Temperature dependence of Fermi momenta and velocities along high-symmetry directions. For the legend
        of designations, see Fig.\,\ref{Fig:Map276K}. Here $k_i$ and $v_i$ denote Fermi momenta and velocities respectively,
        while $d_{ij}$ denotes the distance between points $i$ and $j$ in momentum space. The vertical dotted lines mark
        the temperatures of the commensurate and incommensurate CDW transitions. The dashed lines are drawn as guides
        for the eye. Note that the $\Gamma$-barrel does not change perceptibly in the normal state, whereas both K- and
        M-centered barrels show detectable changes even above $T_{\rm ICDW}$, which are often reversed
        at $T_{\rm ICDW}$, becoming nonmonotonic.}\label{Fig:TdepFS}
\end{figure*}

\begin{table*}[t]
        \begin{tabular}[c]{l|@{~~}c@{~~}c@{~~}c@{~~}c@{~~}c@{~~}c|@{~}c@{~~}c@{~~}c@{~~}c@{~~}c@{~~}c@{}l}
         \toprule
         &&\multicolumn{4}{c}{$\mathrm{\Gamma}$- and K-barrels} &&& \multicolumn{4}{c}{M-barrels (``dogbones'')}&&\\
         $T$\,(K)& $t_0^{\rm a}$ & $t_1^{\rm a}$ & $t_2^{\rm a}$ & $t_3^{\rm a}$ & $t_4^{\rm a}$ & $t_5^{\rm a}$ & $t_0^{\rm b}$ & $t_1^{\rm b}$ & $t_2^{\rm b}$ & $t_3^{\rm b}$ & $t_4^{\rm b}$ & $t_5^{\rm b}$&(eV)\\
         \midrule
         300 & --0.107 & 0.099 & 0.175 & --0.011 & --0.011 &--0.027~ & ~0.321 & 0.151 & 0.368 & 0.010 & --0.024 & 0.046&\\
         200 & --0.110 & 0.088 & 0.178 & --0.015 & --0.013 &--0.039~ & ~0.328 & 0.121 & 0.370 & 0.002 & --0.017 & 0.039&\smash{~$\Bigg{\}}$ normal state\hspace{-20em}}\\
         150 & --0.120 & 0.081 & 0.183 & --0.024 & --0.013 &--0.051~ & ~0.330 & 0.125 & 0.367 & 0.008 & --0.015 & 0.044&\\
         100 & --0.126 & 0.080 & 0.177 & --0.032 & --0.010 &--0.051~ & ~0.364 & 0.168 & 0.385 & 0.035 & --0.008 & 0.082&\smash{~~$\leftarrow$ ICDW state\hspace{-20em}}\\
         \bottomrule
         \end{tabular}
         \caption{Experimental tight-binding parameters of 2H-TaSe$_2$ at different temperatures determined from the
         data shown in Fig\,\ref{Fig:TdepFS}.}
         \label{Table:TaSe2TB}
\end{table*}

The ARPES measurements presented here were performed using 50\,eV photons on a freshly cleaved surface of a 2H-TaSe$_2$ single crystal in an ultra high vacuum of $\sim$\,10$^{-10}$\,mbar. The data were collected within $\sim$\,24 hours after cleavage, whilst no signs of surface degradation have been observed in the ARPES data. The position of the Fermi level has been calibrated by measuring a silver film freshly evaporated on the sample holder before the experiment.

We have investigated the temperature evolution of the 2H-TaSe$_2$ FS in detail by measuring the Fermi momenta and
velocities along all high-symmetry directions as a function of temperature, as shown in Fig.\,\ref{Fig:TdepFS}. All
values were obtained by fitting momentum distribution curves in a finite energy window enclosing the Fermi energy, with
subsequent linear fitting of the resulting experimental dispersion. Though the momentum resolution of our ARPES data was
$\sim$\,0.01\,\AA$^{-1}$, the actual position of the peak in each dataset could be determined with much greater accuracy
\cite{EvtushinskyKordyuk06}, therefore the error bars in Fig.\,\ref{Fig:TdepFS} account both for the momentum resolution
and for uncertainty in the $\mathbf{k}$-space positioning of the ARPES spectra, the latter providing the dominant
contribution to the error. We have observed similar nonmonotonic changes of either Fermi momentum or velocity in several
points along the FS, as seen in panels (d), (h), (k), (m), and (o). At some other points, no temperature dependence
above the ICDW transition is observed, whereas strong changes occur between $T_{\rm ICDW}$ and $T_{\rm CCDW}$ [panels
(b), (g), (j)]. There are no sharp changes at either phase transition, but rather a gradual evolution happening across
the whole temperature range of the ICDW phase. This suggests that the incommensurate state can be interpreted as an
intermediary between the normal and CCDW states, strongly affected by the fluctuations of the commensurate phase
\cite{LeeRice73}.

\begin{figure}[b]
        \includegraphics[width=\columnwidth]{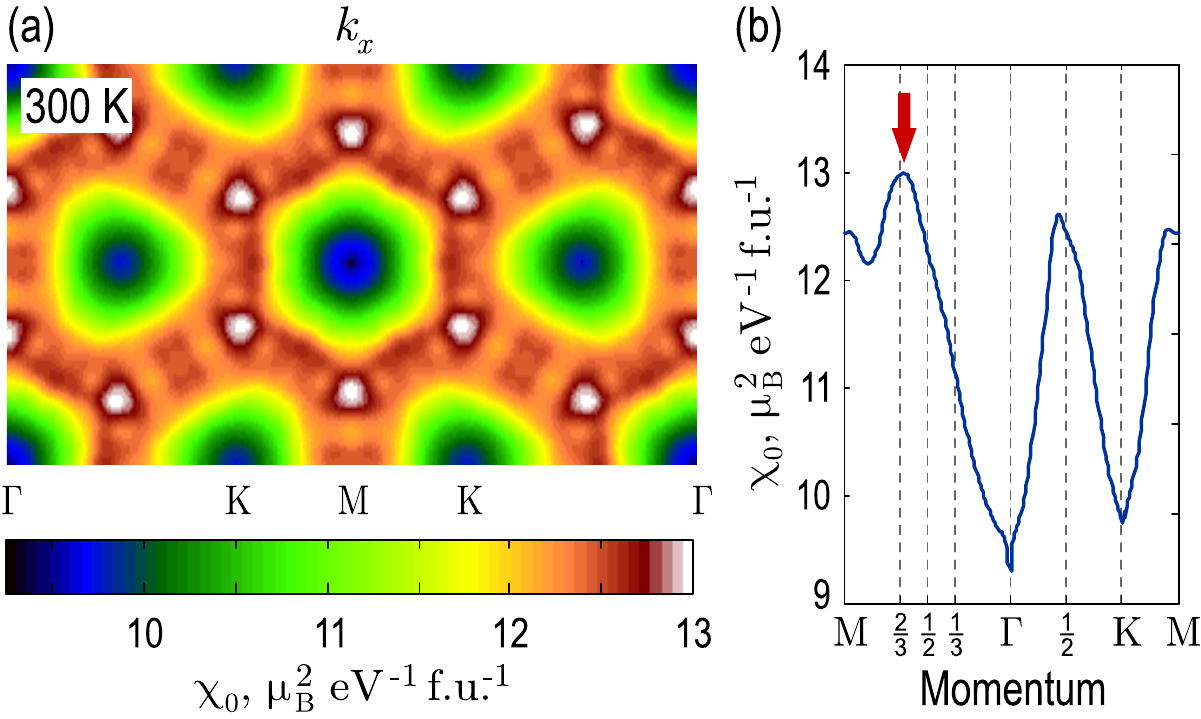}
        \caption{Nesting properties of the room-temperature Fermi surface of 2H-TaSe$_2$. (\textbf{a})~Color map of
        the Lindhard function in the static limit ($\omega=0$) calculated at 300\,K as a function of $k_x$ and $k_y$. White areas correspond to the dominant nesting vectors. (\textbf{b})~Profile of the Lindhard function along high-symmetry directions. The arrow marks the
        position of the nesting vector.\vspace{-4pt}}\label{Fig:Nesting}
\end{figure}

The present measurements enabled us to refine the tight-binding parameters of the electronic structure to a greater
accuracy than was previously achievable \cite{InosovZabolotnyy08}. By a procedure similar to that of
Ref.\,\onlinecite{InosovZabolotnyy08}, we have fitted the experimental dispersion to the data using the following
tight-binding expansion (quasimomenta $k_x$ and $k_y$ enter the formula in dimensionless units):
\begin{equation*}
\begin{split}
\epsilon_\mathbf{k} = t_0 &+
t_1\Bigl[2\,\mathrm{cos}\,\frac{k_x}{2}\,\mathrm{cos}\,\frac{\sqrt{3}\,k_y}{2}+\mathrm{cos}\,k_x\Bigr]\\ &+
t_2\Bigl[2\,\mathrm{cos}\,\frac{3k_x}{2}\,\mathrm{cos}\,\frac{\sqrt{3}\,k_y}{2}+\mathrm{cos}\,\sqrt{3}\,k_y\Bigr] \\&+
t_3\,[\,2\,\mathrm{cos}\,k_x\,\mathrm{cos}\,\sqrt{3}\,k_y+\mathrm{cos}\,2k_x]\\ &+
t_4\Bigl[2\,\mathrm{cos}\,\frac{3k_x}{2}\,\mathrm{cos}\,\frac{3\sqrt{3}\,k_y}{2}+\mathrm{cos}\,3\,k_x\Bigr] \\&+
t_5\Bigl[\mathrm{cos}\,\frac{k_x}{2}\,\mathrm{cos}\,\frac{3\sqrt{3}\,k_y}{2}+\mathrm{cos}\,\frac{5k_x}{2}\,\mathrm{cos}\,\frac{\sqrt{3}\,k_y}{2}\\&
\qquad\qquad\qquad\qquad\qquad\quad\quad~+\mathrm{cos}\,2k_x\,\mathrm{cos}\,\sqrt{3}\,k_y\Bigr].
\end{split}
\end{equation*}
The tight-binding parameters were independent for the two bands, which resulted in the total of 12 fitting parameters.
These were then found by solving an overdetermined system of 16 equations relating the Fermi momenta and velocities of
the model to the experimentally measured ones, so that the resulting tight-binding model best reproduces both the FS contours and the experimental dispersion in the vicinity of the Fermi level. The corresponding tight-binding
parameters are given in Table~\ref{Table:TaSe2TB} for several temperatures. With improved accuracy, we can confirm our
conclusions made in Ref.\,\onlinecite{InosovZabolotnyy08}, namely that there is a strong nesting vector of the normal-state
FS in the vicinity of the CDW wave vector, as determined from the Lindhard function calculated from
experimental dispersions \cite{InosovZabolotnyy08, InosovBorisenko07} (see Fig.\,\ref{Fig:Nesting}). The
incommensurability of the nesting peak appears to be smaller than we have previously estimated
\cite{InosovZabolotnyy08}, and accounts to only 3$\pm$1\% of the nesting vector, which is in good agreement with
the experimentally observed incommensurability parameter of 2\% \cite{MonctonAxe75}.

\begin{figure}[t]
        \includegraphics[width=0.7\columnwidth]{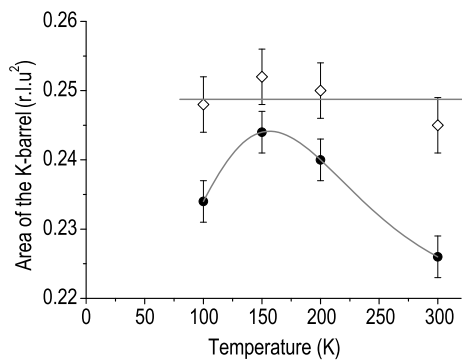}
        \caption{The effective area of the K-centered barrel as a function of temperature with ($\diamond$) and without
        ($\bullet$) the pseudogap correction.\vspace{-4pt}}\label{Fig:KBarrel}
\end{figure}

Using the simple model of the PG introduced in Ref.\,\onlinecite{EvtushinskyKordyuk08} and its temperature dependence from
Ref.\,\onlinecite{BorisenkoKordyuk08}, we could estimate its effect on the remnant FS and compare it with our tight-binding
models. The dominant contribution to the change in FS area comes from the K-barrel, where the PG is largest. In
Fig.\,\ref{Fig:KBarrel}, the area of the K-barrel is plotted as a function of temperature ($\bullet$) to show its
nonmonotonic behavior, which has almost vanished after we accounted for the effective increase of the barrel due to the
opening of the PG ($\diamond$). This confirms that the evolution of FS contours is caused by the PG and provides an
independent piece of evidence for the opening of the PG already in the normal state, far above the ICDW transition
temperature. The doping level estimated from the FS area at 300\,K proved to be 2.006$\pm$0.01 electrons per unit cell
in perfect agreement with the nominal doping.

\vspace{-5pt}\section{Summary and discussion}\vspace{-5pt}

To summarize, in the present paper we have presented evidence for the non-monotonic temperature dependence of the remnant Fermi surface in 2H-TaSe$_2$, which we associate with the similar behavior of the pseudogap, in a wide temperature range from 20\,K to the room temperature, and in different regions of the reciprocal space. This became possible after the high-accuracy determination of the FS geometry using a newly proposed method, based on the measurements of electron momenta and velocities along all high-symmetry directions of the reciprocal space as functions of temperature with subsequent fitting to a tight-binding model. Comparison of the calculated nesting vector with the known propagation vector of the CDW order, as well as the evaluation of the electron doping from the Fermi surface area, indicate the highest accuracy of the resulting band structure.

Let us now discuss the possible reasons for the anomalous nonmonotonic behavior of the Fermi momenta, velocities, and PG in the vicinity of the ICDW transition. The possible origin of the gradual changes in these parameters observed between $T_{\rm ICDW}$ and $T_{\rm CCDW}$ is the competition between incommensurate order and CCDW fluctuations. As known from X-ray scattering measurements \cite{ClancyGaulin07}, in the simplest possible density wave systems\,---\,quasi-one-dimensional Heisenberg chain materials\,---\,the incommensurate order and commensurate order fluctuations coexist in the PG phase and compete against each other in the temperature range between the two phase transitions. Moreover, the commensurate fluctuations in these model materials behave nonmonotonically in the vicinity of the incommensurate phase transition temperature (Ref.\,\onlinecite{ClancyGaulin07}, Fig.\,3~and~4). A similar effect is also observed in three-dimensional metal alloys \cite{EcclestonHagen96}. This is in line with our observations of the nonmonotonic changes of the FS shape and the value of the PG, confirming the hypothesis that fluctuations of the CDW order could be responsible for these effects. Obtaining more direct evidence that might confirm our hypothesis would require a similar measurement of the strength of CDW fluctuations as a function of temperature in 2H-TaSe$_2$ by an independent momentum-resolved technique such as X-ray scattering, which to our best knowledge still awaits to be performed.

\makeatletter\renewcommand{\fnum@figure}[1]{\figurename~\thefigure.}\makeatother

\vspace{-5pt}\section{Acknowledgements}\vspace{-5pt}

This~project~is~part~of~the~Forschergruppe~FOR538 and is supported by the DFG under Grants No.~KN393/4 and BO1912/2-1.
The work in Lausanne was suppor\-ted by the Swiss National Science Foundation and by the MaNEP. ARPES experiments were
performed using the $1^3$ ARPES end station at the UE112-lowE PGMa beamline of the Berliner
Elektronenspeicherring-Gesellschaft für Synchrotron Strahlung m.b.H. (BESSY). We thank R.\,Hübel for technical support.

\end{document}